\documentstyle[12pt,aaspp4,epsf]{article}
\input epsf
\input{amssym.def}      %
\input{amssym.tex}      
\def\kms{{\rm km/s}}

\def\mum{\mu {\rm m}}

\def\hpc{{\rm h}^{-1}\,{\rm pc}}
\def\lsun{{\,L_\odot}}
\def\g0{{\,{\rm G}_0}}
\def\lcii{{\rm F_{\rm [CII]}}}
\def\l2cii{{\rm L_{\rm [CII]}}}
\def\ffir{{\rm F_{\rm FIR}}}
\def\lfir{{\rm L_{\rm FIR}}}
\def\rat{{\rm L_{\rm [CII]}}/{\rm L_{\rm FIR}}}

\def\lb{{\rm L_{\rm B}}}
\def\6f{{\rm F_\nu(60 \mum)}}
\def\f100{{\rm F_\nu(100 \mum)}}

\def\@journalname{Astrophysical Journal}
\def\accepted#1{\gdef\@accptdate{#1}}
\accepted{\relax}
\def\journalid#1#2{\gdef\@jourvol{#1}\gdef\@jourdate{#2}}
\def\articleid#1#2{\gdef\@startpage{#1}\gdef\@finishpage{#2}}
\begin{document}

\title{ISO measurements of [CII] line variations in galaxies}
\author{S. Malhotra, \altaffilmark{1} 
G. Helou, \altaffilmark{1} 
G. Stacey, \altaffilmark{2} 
D. Hollenbach, \altaffilmark{3} 
S. Lord, \altaffilmark{1} 
C. A. Beichman \altaffilmark{1} 
H. Dinerstein \altaffilmark{4}  
D. A. Hunter \altaffilmark{5}  
K. Y. Lo \altaffilmark{6}  
N. Y. Lu \altaffilmark{1} 
R. H. Rubin \altaffilmark{3}
N. Silbermann \altaffilmark{1} 
H. A. Thronson Jr. \altaffilmark{7} 
M. W. Werner \altaffilmark{8}
}

\begin{abstract}

We report measurements of the [CII] fine structure line at 157.714
$\mum$ in 30 normal star-forming galaxies with the Long Wavelength
Spectrometer (LWS) on the Infrared Space Observatory (ISO). The ratio
of the line to total far-infrared luminosity, ($\rat$), measures the
ratio of the cooling of gas to that of dust; and thus the efficiency
of the grain photoelectric heating process. This ratio varies by more
than a factor of 40 in the current sample. About two-thirds of the
galaxies have $\rat$ ratios in the narrow range of $2-7\times
10^{-3}$. The other one-third show trends of decreasing $\rat$ with
increasing dust temperature, as measured by the flux ratio of infrared
emission at 60 and 100 $\mum$ ($\6f/\f100$); and with increasing
star-formation activity, measured by the ratio of far-infrared and
blue band luminosity $\lfir/\lb$.  We also find three FIR bright
galaxies which are deficient in the [CII] line, which is undetected
with $3\sigma$ upper limits of $\rat < 0.5-2\times10^{-4}$.

The trend in the $\rat$ ratio with the temperature of dust and with
star-formation activity may be due to decreased efficiency of
photoelectric heating of gas at high UV radiation intensity as dust
grains become positively charged, decreasing the yield and the energy
of the photoelectrons. The three galaxies with no observed PDR lines
have among the highest $\lfir/\lb$ and $\6f/\f100$ ratios.  Their lack
of [CII] lines may be due to a continuing trend of decreasing $\rat$
with increasing star-formation activity and dust temperature seen in
one-third of the sample with warm IRAS colors. In that case the upper
limits on $\rat $ imply a ratio of UV flux to gas density $\g0/n > 10$
cm$^3$ (where $\g0$ is in the units of local average interstellar
field). The low $\rat$ could also be due to either weak [CII] because
of self-absorption or strong FIR continuum from regions weak in [CII],
such as dense HII regions or plasma ionized by hard radiation of
AGNs. The mid-infrared and radio images of these galaxies show that
most of the emission comes from a compact nucleus. CO and HI are
detected in these galaxies, with HI seen in absorption towards the
nucleus.

\keywords{radiation mechanisms: thermal, ISM: atoms,galaxies: ISM, infrared: ISM: lines and bands }
\end{abstract}

\altaffiltext{1}{IPAC, 100-22, California Institute of Technology, Pasadena,
CA 91125} 
\altaffiltext{2}{Cornell University,  Astronomy Department, 220 Space Science
Building, Ithaca, NY 14853}
\altaffiltext{3}{NASA/Ames Research Center, MS 245-3, Moffett Field, CA 94035}
\altaffiltext{4}{University of Texas, Astronomy Department, RLM 15.308, Texas, Austin, TX 78712}
\altaffiltext{5}{Lowell Observatory, 1400 Mars Hill Rd., Flagstaff, AZ 86001}
\altaffiltext{6}{University of Illinois, Astronomy Department, 1002 W. Green St., Urbana, IL 61801}
\altaffiltext{7}{University of Wyoming, Wyoming Infrared Observatory, Laramie, WY, 82071}
\altaffiltext{8}{Jet Propulsion Laboratory, MS 233-303, 4800 Oak Grove Rd., Pasadena, CA 91109}

\section{Introduction}
The $C^{+}$ fine structure transition at $157.714 \mum$ is an
important coolant of the warm neutral interstellar medium. This is
because carbon is the fourth most abundant element and has a lower
ionization potential (11.26 eV) than hydrogen, so that carbon will be
in the form of $C^{+}$ on the surfaces of far-UV illuminated neutral
gas clouds.  The depth of these photodissociation regions (PDRs) is
determined by dust extinction, which is typically $A_v \le
4$. Secondly, the $158 \mum$ [CII] line is relatively easy to excite
($\Delta$E/k $\simeq 91$ K), so that $C^{+}$ can cool warm neutral gas
where the two most abundant atoms H and He cannot (cf. Tielens \&
Hollenbach (1985), hereafter TH85, Wolfire, Tielens \& Hollenbach
(1990), hereafter WTH90). Gas heating in PDRs is dominated by
energetic photoelectrons ejected from dust grains following UV photon
absorption (Watson 1972).

The [CII] line is roughly 1500 times more intense than the CO($1\to
0$) rotational line in normal galaxies and Galactic molecular clouds;
and 6300 times more intense than the CO($1\to 0$) line in starburst
nuclei and Galactic star formation regions (Crawford et al., 1985,
Stacey et al. 1991).  For normal galaxy nuclei and star formation
regions in the spiral arms, most of the observed [CII] arises from
PDRs on molecular cloud surfaces.  However, integrated over the disks
of normal spiral galaxies, a substantial fraction (up to 50\%) may
also arise from "standard" atomic clouds, i.e. the cold neutral medium
(CNM) (Madden et al. 1993) or from extended, low density HII regions
(Heiles 1994).

With the Long Wavelength Spectrometer (LWS, Clegg et al. 1996) on the
Infrared Space Observatory (ISO, Kessler et al. 1996) it is now
possible to measure the [CII] line emission from normal, not very
infrared-bright galaxies (Lord et al. 1996). Observations of fine
structure atomic and ionic lines in the far-infrared have been
obtained for half the sample of about 60 galaxies in the US-ISO Key
Project on Normal Galaxies. The galaxies were selected to span a range
of morphologies: Irr-Sa in this paper, and Irr-E in the complete
sample (see Helou et al. 1996 for sample selection criteria), dust
temperature (as indicated by the ratio of flux density at $60 \mum$
and $100 \mum$), and star-formation activity (measured by the ratio of
FIR to Blue luminosity). This paper offers a first look at statistical
behavior of cooling via [CII] line in a diverse sample of galaxies,
aimed at a better understanding of the physical conditions in the ISM
of star-forming galaxies.

\section{Observations}

With the $80 \arcsec$ beam of LWS and spectral resolution of $0.6
\mum$ we measure total line flux for the present sample of
galaxies. The galaxies were selected to have $FWHM < 0.5\arcmin$ in
FIR emission using deconvolved IRAS maps. LWS measures the total flux
and not the brightness temperature so beam dilution is irrelevant. The
[CII] line observations were planned to achieve ($1\sigma$)
sensitivities of $5\times 10^{-5}\times \ffir$, where $\ffir$ is the
total far-infrared flux of the galaxy and is computed according to the
formula $ \ffir=1.26 \times 10^{-14} [2.58 \times F_{\nu}(60\mum) +
F_{\nu}(100\mum)] W m^{-2}$ (Helou et al. 1988), where
$F_{\nu}(60\mum)$ and $F_{\nu}(100\mum)$ are flux densities in Jansky
at 60 and 100 $\mum$. For comparison, previous observations of Milky
Way, starburst galaxies and galactic nuclei show that the line to
continuum luminosity ratio $\rat = 1-10\times 10^{-3}$ (Stacey et
al. 1985, Wright et al. 1991, Crawford et al. 1985, and Stacey et
al. 1991, Lord et al. 1996). In Galactic PDR sources associated with
HII regions the $\rat$ ratio varies from $\sim 3\times 10^{-3}$ in
PDRs (e.g. NGC 2023) to $\sim 8\times 10^{-5}$, decreasing with HII
region density and UV flux (cf. Crawford et al. 1985, Hollenbach,
Takahashi \& Tielens 1991).

The data were reduced and calibrated with the ISO pipelines OLP4.2 to
OLP5.2. The line profiles were derived from several scans by running a
median boxcar filter through typically six spectral scans per
object. We use the median of the observed fluxes instead of the mean
to reduce the influence of outlying points arising from cosmic ray
hits. The flux in the [CII] line was determined by directly
integrating under the line. The upper limits on non-detections were
derived by calculating the flux from a hypothetical gaussian line with
an amplitude of $3\sigma$ and the effective instrumental profile,
since the line is unresolved for all sources.

The line fluxes suffer from calibration uncertainties mainly due to
uncorrected fluctuations of detector responsivities with time. To
estimate the calibration errors we compared the observed continuum
levels at $158 \mum$ with flux predicted from dust emission models
made using IRAS 100 and 60 $\mum$ fluxes (Helou \& Beichman 1990).
This comparison shows that the mean ISO/IRAS continuum ratio at $158
\mum$ is about 1.1 and the observed fluxes lie within a factor of two
of the models.  These values are conservative estimates of the
uncertainty in the line fluxes because the continuum levels, which are
compared with IRAS flux densities, are affected by
ill-determined dark currents (additive in nature) as well as detector
responsivities (multiplicative), whereas the line fluxes are affected
only by the detector responsivities. Thus we are confident of the
trends in Figure 1 which span a factor of $\simeq 40$ in the $\rat$
ratio.

\section{Results and Discussion} 

The ratio of [CII] line to FIR continuum flux from neutral gas depends
on the efficiency of heating of gas by photoelectrons from dust, and
is expected to be sensitive to the physical conditions in the ISM. We
therefore explore the dependence of $\rat$ on the dust temperature
(characterized by $\6f/\f100$) and global star-formation activity in
galaxies (characterized by $\lfir/\lb$).  The two quantities,
$\lfir/\lb$ and $\6f/\f100$ are positively correlated with each other,
i.e. galaxies with more active star-formation also have higher dust
temperatures (Soifer \& Neugebauer 1991). The present sample does not
reflect this correlation as strongly as a randomly selected sample, as
the galaxies were selected to span the parameter space in $\lfir/\lb$
and $\6f/\f100$. In Figure 1 the quantity log($\rat$) is plotted as a
function of the ratio of FIR and B-band luminosity $\lfir/\lb$ of the
galaxies, and as a function of flux density ratio at 60 and 100 $\mum$
($\6f/\f100$). Three main results emerge from Figure 1, and are
discussed in the next three subsections.

\begin{figure}[htb]
\setbox3=\vbox{\epsfxsize=3.3in
\epsfbox{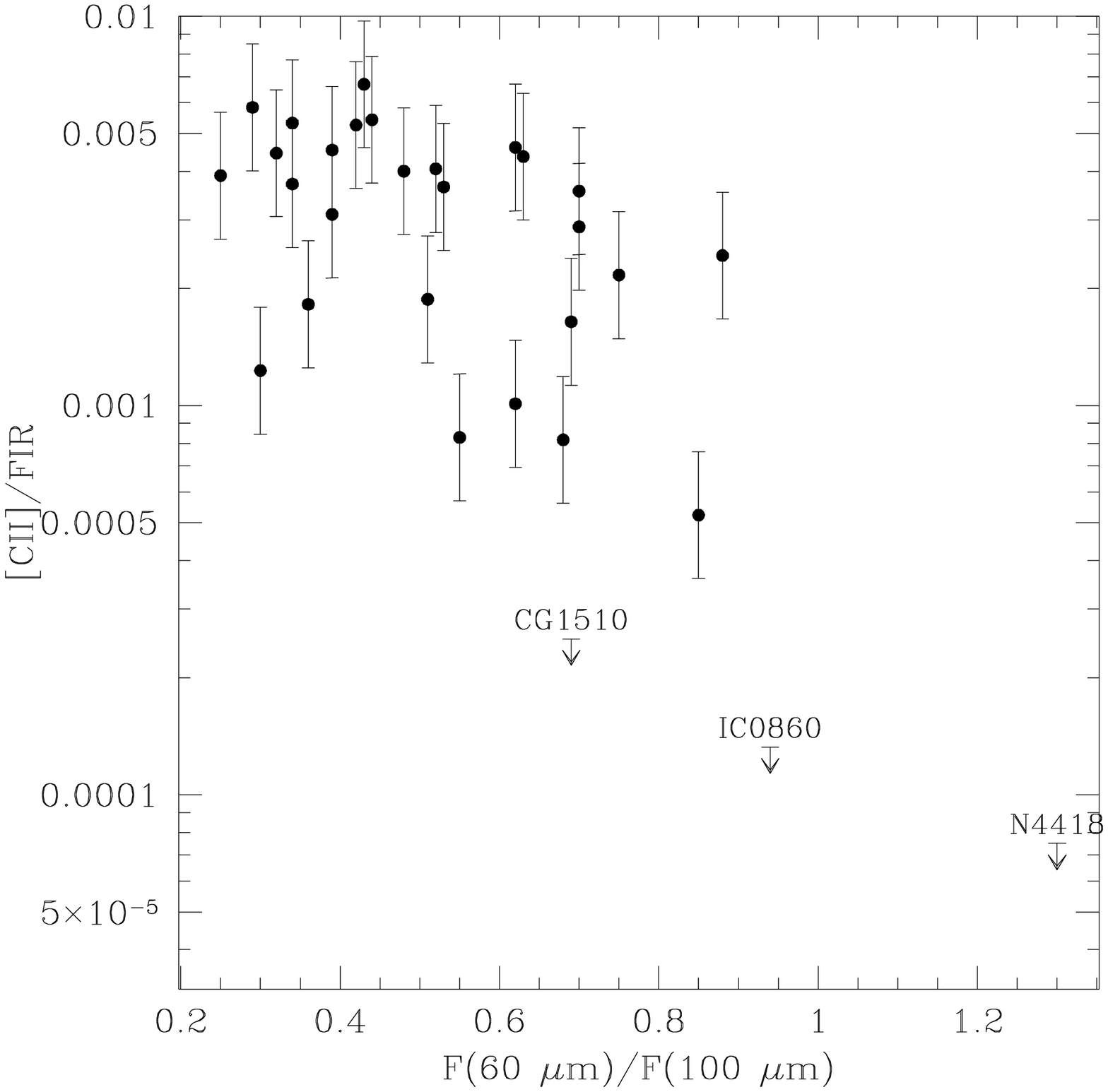}}
\centerline{\box3
\epsfxsize=3.3in \epsfbox{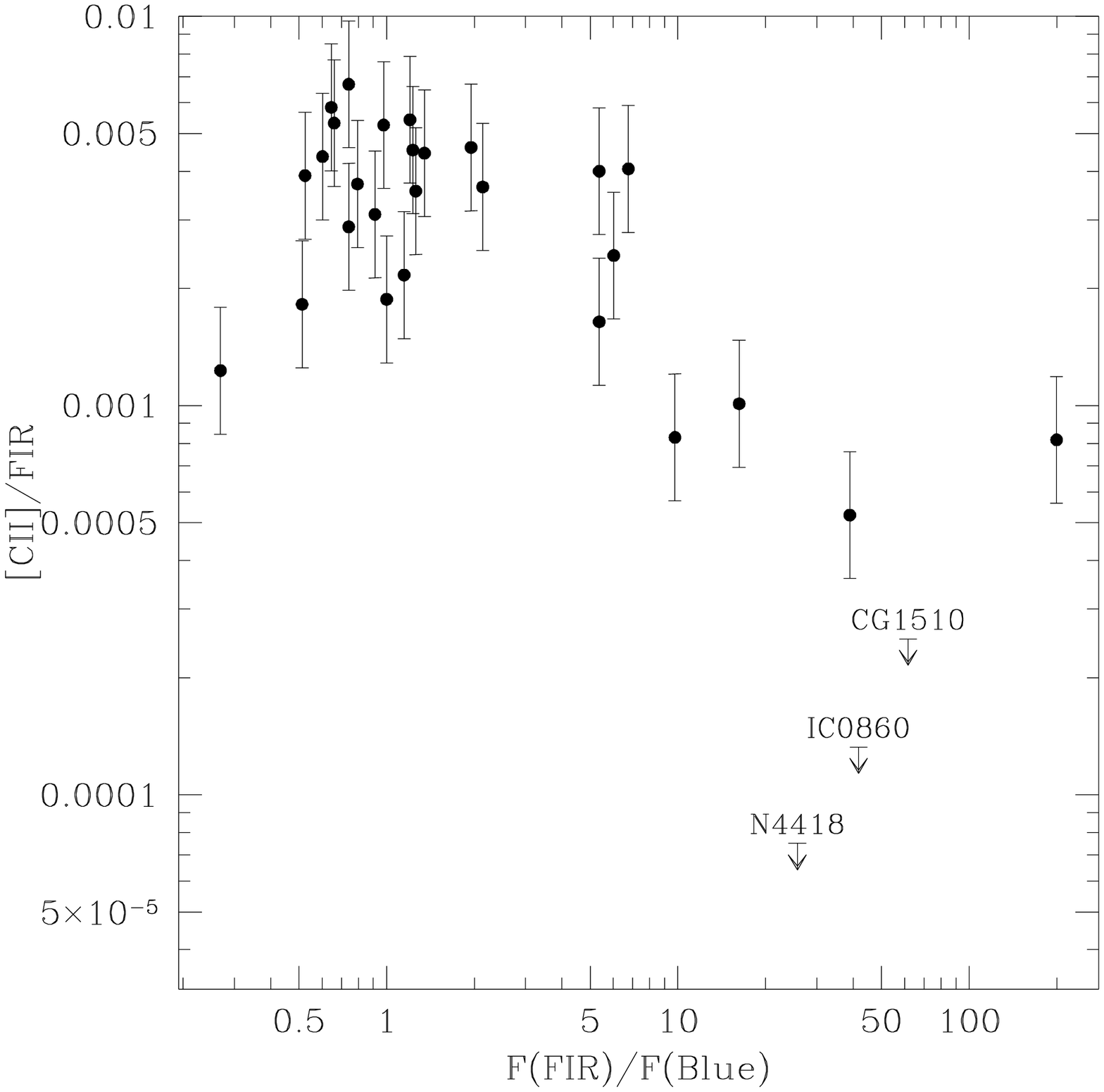} }
\caption{ The ratio of FIR and [CII] line luminosities is plotted
against the dust temperature characterized by the ratio of flux
densities at $60 \mum$ and $100 \mum$ from IRAS measurements. The
error bars indicate the dominant source of uncertainty, namely the
calibration uncertainty of a factor of two. There is a trend for
galaxies with higher $\6f/\f100$ (warmer dust) to have lower
$\rat$. Three galaxies in a sample of 30 have no detected [CII], they
are identified and shown as upper limit symbols in the figure. (b)
$\rat$ shows a similar trend with the ratio of FIR/Blue luminosity,
which can be used as an indicator of star formation activity. Galaxies
with higher $\lfir/\lb$ (more active star formation) have lower $\rat$.}
\end{figure}

\subsection{The modal value of $\rat$:} 

About two-thirds of the galaxies (19 out of 30) have $\rat$ ratio
between 0.2-0.7\% (Figure 1).  This is consistent with the value
measured for the Milky Way, and both normal and starburst galaxies
(Stacey et al. 1985, Crawford et al. 1985, Wright et al. 1991, Stacey
et al. 1991, Madden et al.  1993, Lord et al. 1996). These observed
ratios are well understood in terms of PDR models (TH85, WTH90). The
[CII] line originates from warm neutral gas in the photodissociation
regions. The efficiency of the photoelectric heating of gas is a
fraction of a percent. In the data presented in this paper we see a
confirmation of this model in galaxies spanning a range of
morphologies, with dust temperatures and $\lfir/\lb$ ratios which are
not unusually high.

\subsection{Trends in the measured $\rat$:} 

About 1/3 of the present sample with higher FIR/Blue luminosity and
warmer far-infrared colors (i.e. higher $\6f/\f100$ ratio) show a trend of
decreasing $\rat$ with $\6f/\f100$ and $\lfir/\lb$ (Figure 1). Such a
trend is also seen within our Galaxy (Nakagawa et al. 1995, Bennett et
al. 1994).  The trends in $\rat$ with $\lfir/\lb$ and with $\6f/\f100$
could be due to the following. (Any reference to the trends refers
only to the one-third of the sample with warm FIR colors which shows
decreasing $\rat$ with $\lfir/\lb$ and with $\6f/\f100$)

(a) For high ratios of UV flux to gas density ($\g0/n$), the dust
grains are positively charged and are less efficient at heating the
gas because the potential barrier to photoejection is higher
(TH85). This leads to smaller PDR line to dust continuum ratios
because, most of the UV energy is absorbed and reradiated in the FIR
by grains, and a smaller fraction of the UV energy goes into gas
heating, so the gas requires less cooling.

(b) At densities of $n > 10^4 cm^{-3}$ $C^+$ ions become collisionally
de-excited (TH85, WTH90) and an increasing amount of cooling is done
through other channels; for example, the [OI] line at $63 \mum$. We
see that this line cannot compensate completely for the decrease of
$\rat$ in Figure 1, as the observed [OI]($63 \mum$) line flux does not
exceed $3\times$[CII] in this sample, whereas the ratio $\rat$ varies
by a factor of 40. CO and CI lines can also cool the gas.

(c) In each galaxy we measure the total FIR and the total [CII] line
emission arising from a collection of HII regions, PDRs, and diffuse
atomic clouds. Galactic HII regions like W~51 and S~140 have smaller
$\rat$ ratios compared to typical PDRs (Emery et al. 1996). A varying
mix of these components will lead to variations in the $\rat$ ratio
seen in Figure 1.

\subsection {[CII] deficient galaxies:}

In the extreme cases of NGC~4418, IC~0860 and CGCG~1510.8+0725, 
[CII] emission is undetected at the $3\sigma$ level of $\l2cii =5-0.5
\times 10^{-4} \lfir$. Emission in the other prominent PDR cooling
line [OI] at $63 \mum$ is undetected at similar levels, as are [OIII]
($88 \mum$) and [NII] ($122 \mum$). In Figure 1 we show $3\sigma$
upper limits on the flux in the [CII] $158 \mum$ transition.

The absence of [CII] emission in these galaxies is very striking
because the upper limits on the $\rat$ in these galaxies are up to 40
times smaller than the narrow range in $\rat$ measured for two-thirds
of the sample. Yet these galaxies have normal atomic and molecular gas
components. A comparable upper limit on $\rat$ was reported by Stacey
et al. (1991) for the IR luminous galaxy NGC~660. Luhman et al. (1997)
also report on weak [CII] emission from ultraluminous galaxies.

In order to understand the absence of the [CII] cooling line, let us
review what is distinct about these three galaxies. They have among
the highest $ \lfir/\lb $ luminosity ratio and dust temperature
$\6f/\f100$ in the sample, but are not ultraluminous IR sources. Their
FIR luminosities vary from Log($L_{FIR}/\lsun$)=10.05-10.24. The
mid-infrared images at 7 and 15 $\mum$ are unresolved at $7 \arcsec$
resolution of ISOCAM (Silbermann et al. 1997) and they are compact in
radio continuum at 1.4 GHz (Condon et al 1990): NGC~4418 measures
$0.5'' \times 0.3''(32 \hpc)$, CGCG~1510.8+0725 measures $0.5'' \times
0.5'' (85 \hpc)$ and IC~860 $0.4'' \times 0.3 '' (75 \hpc)$. Two other
galaxies in the present sample, NGC~3620 and IRAS~F10565+2448, have
comparable FIR/Blue luminosity and $\6f/\f100$ but do show weak [CII]
emission lines. NGC~3620 is not compact in mid-IR emission and
IRAS~F10565+2448 is a point source at high redshift in the sample, so
the size limits are less stringent (the size $< 5.5$ kpc).
CGCG~1510.8+0725 shows a broad H$\alpha$ line indicative of a Seyfert
nucleus (Kim et al. 1995) but no H$\beta$, (possibly due to
obscuration), causing it to be classified as a HII galaxy (Veilleux et
al. 1995). NGC~4418 shows many indications of being a hidden Seyfert
nucleus (Roche et al. 1986, Kawara, Nishida \& Philips 1989), and has
the distinction of having the most 'extinguished' nucleus with $A_V >
50$ (Roche et al. 1986) as seen in the $9.7 \mum$ absorption feature.

There are several possible explanations for the [CII] deficiency in
these three systems.

(A) For high ratios of UV radiation to density of the warm gas
($\g0/n$), the dust grains are positively charged and are less
efficient at heating the gas (TH85). This is also a likely explanation
for the trend of decreasing $\rat$ discussed in section
3.2(a). Wolfire et al. (1990) predict $(\lcii+\rm F_{\rm[OI]})/\ffir <
10^{-4}$ for $\g0/n > 10$ cm$^3$.  A high $\g0/n$ can come about in low
density or high radiation regions. CO observations of these galaxies
show copious, compact, circumnuclear gas (Planesas, Mirabel \& Sanders
1991, Kwara et al. 1990). This indicates that even though the density
of the gas $n$ is high, $\g0/n$ is higher.

(B) The addition of a compact FIR source in the nucleus of galaxies, e.g. an
obscured AGN, can lead to both a higher FIR/Blue luminosity ratio and
a lower $\rat$ ratio. This possibility is supported by the compact
appearance of the mid-IR and radio continuum emission in these three
galaxies. A buried AGN could produce a lot of FIR without much [CII]
since the UV field will be inefficient at making [CII] both due to its
hardness (higher ionization states for C will be common) and due to
its strength (inefficiency of heating gas as discussed in point (A)).

(C) The [CII] line may be optically thick or the lines may be weak due
to self-absorption by lower excitation $C^{+}$ foreground gas. Such
self-absorption is seen in the $158 \mum$ line observed at high
resolution (Boreiko \& Betz 1997). This hypothesis is supported by the
compactness of the mid-IR emission in the three galaxies. The [CII]
line becomes optically thick at $N(C^{+})=5 \times 10^{17} $cm$^{-2}$
for a velocity width of 4 $\kms$ (Russell et al. 1980). Assuming a
velocity width of $120 \kms$ for NGC 4418 (Sanders, Scoville \& Soifer
1991) and $N(C^{+})/N(H) \simeq 3.7 \times 10^{-4}$, optical depth of
one in the [CII] line is reached for column densities of $N(H)=4
\times 10 ^{22}$ cm$^{-2}$.  Roche et al. (1986) estimate the visual
extinction towards the nucleus of NGC~4418 to be $120 > A_V > 50$. If
the PDR gas has $\tau=1$ it will contribute $A_V \simeq 20$.  This
high a PDR contribution to the extinction would imply a high ratio of
PDR to cold molecular gas, instead of the conventional picture where
PDRs form a thin surface layer of molecular clouds. Since atomic
oxygen is found in abundance in dark clouds the non-detection of the
[OI] $63 \mum$ line could easily be due to self-absorption as well
(Poglitsch et al. 1996, Kraemer, Jackson \& Lane, 1996). Compact
concentrations of circumnuclear molecular gas in CGCG~1510.8+0725 and
NGC~4418 (Planesas, Mirabel \& Sanders 1991, Kwara et al.1990) further
support the scenario where the CII line is at high optical depth or
self-absorbed.

(D) The dominant PDR cooling lines [CII] and [OI] could be weak or
missing if most of the FIR emission is from dust in high density HII
regions (c.f. point (c) in section 3.2). This hypothesis is disfavored
because the dust temperatures derived from the continuum shapes are
too low compared to HII regions in the Milky Way (e.g. S140, Emery et
al. 1996).

(E) An aging starburst, and a weak UV field relative to a softer,
optical/infrared field which heats the dust, is used to explain the
[CII] line weakness in the Galactic Center (Nakagawa et al. 1995).
This mechanism seems unlikely because the [CII] deficiency is
systematically found in galaxies with warmer FIR colors and higher
$\lfir/\lb$ ratios indicating younger starbursts, with the hotter dust
pointing to the presence of massive stars and UV photons.

\section{Summary and Conclusions}

Three main results emerge in this study of [CII] fine structure
transition at $158 \mum$, a primary coolant of photodissociation
regions in star-forming galaxies.

(1) About two-thirds of the observed galaxies show a ratio of $\rat =
0.2-0.7 \%$. This is consistent with the models and efficiencies of
heating of PDR gas by photoelectrons from dust grains illuminated by
moderate UV fluxes.

(2) As the dust temperature and the star-formation rate in galaxies
increase, the ratio of $\rat$ decreases. This is probably due to
higher UV radiation fields in warmer galaxies causing dust grains to
become positively charged, thus leading to a decrease in gas
heating efficiency of photoelectrons from dust grains. This effect is
seen in the warmer one-third of the sample.

(3) In the present sample there are three galaxies near the extreme
end of this trend in the $\6f/\f100$ ratio and in the ratio of
$\lfir/\lb$ which show no detectable line emission in the [CII] and
[OI] lines.  Since the upper limits on $\rat$ for the three galaxies
is 10-40 times smaller than the $\rat$ observed for most galaxies, we
conclude that less than $10\%$ of the FIR emission in these galaxies
arises from the typical PDR gas having moderate UV fluxes.

Upper limits on the [CII] line flux in the three galaxies with no
detection are not very far from the trend of decreasing $\rat$ with
increasing star-formation activity and dust temperature. In our
preferred scenario, both the trend in Figure 1 and the absence of PDR
lines in the extreme galaxies can be naturally explained by a trend of
increasing $\g0/n$ in the hotter, more active galaxies, and hence a
lower efficiency of gas heating. A large contribution to the FIR flux
by an active nucleus or dense HII regions could explain the absence of
[CII] lines but not easily the trends in $\rat$. The other scenarios
where the absence of [CII] line is due to optically thick conditions,
self-absorption or collisional de-excitation are not ruled out by the
present data but other cooling channels for the gas need to be
identified.

The [CII] line at $158 \mum$ has been used as a measure of
star-formation activity (at least of massive stars) (e.g. Stacey et
al. 1991). These ISO-LWS data allow us to probe the variations in the
line strength and how sensitive they are to the physical conditions in
the ISM. If the variations observed in this study are mostly due to
AGN, then the [CII] line strength and its ratio to $L_{FIR}$ is mostly
a diagnostic of whether the FIR luminosity of any given source is due
to star-formation or active nucleus.  However, if the decreasing
$L_{CII}/L_{FIR}$ is caused by increasing $\g0/n$, then these ISO-LWS
data allow us to probe the interplay between star formation and
physical conditions in the ISM.

\acknowledgements
This work was supported by ISO data analysis funding from the US
National Aeronautics and Space Administration, and carried out at the
Infrared Processing and Analysis Center and the Jet Propulsion
Laboratory of the California Institute of Technology. ISO is an ESA
project with instruments funded by ESA Member States (especially the
PI countries: France, Germany, the Netherlands and the United
Kingdom), and with the participation of ISAS and NASA. We thank the
referee for numerous comments.

\end{document}